\documentclass[preprint,12pt,authoryear]{elsarticle}

\usepackage{amssymb}

\usepackage{lineno}
\journal{JHEAP}
\usepackage{amsmath}
\usepackage{siunitx}
\usepackage{threeparttable}
\usepackage{pdfpages}
\usepackage{times}
\usepackage{comment}
\usepackage{rotating}
\usepackage{multirow}
\usepackage{pgfplots}
\usepackage{float}
\usepackage[center,font=small]{caption}
\usepackage[font=scriptsize]{caption}
\usepackage{tikz}
\usetikzlibrary{fit,shapes.geometric}
\usetikzlibrary{plotmarks}
\usepackage{mathdots}
\usepackage{etex}
\usepackage{epsfig}
\usepackage{multirow}
\usepackage{indentfirst}
\usepackage{lipsum}
\usepackage{setspace}
\usepackage{mathtools}
\usepackage{nomencl}
\usepackage{pdfpages}
\DeclareUnicodeCharacter{2212}{-}
\usepackage{graphicx}
\usepackage{subcaption}
\usepackage{rotating}    
\usepackage{longtable}
\usepackage[graphicx]{realboxes}%
\usepackage{emptypage}
\usepackage[inner=35.4mm,outer=25.4mm]{geometry}
\usepackage{gensymb}
\usepackage{hyperref}[colorlinks, citecolor=blue]
\usepackage{multicol}
\usepackage{natbib}

\begin{document}

\begin{frontmatter}

\title{X-ray properties of high-redshift Radio Loud and Radio Quiet Quasars observed by Chandra}

\author[a]{F. Shaban \corref{cor1}}
\ead{fshaban@sci.cu.edu.eg}
\cortext[cor1]{corresponding author}

\author[b]{A. Siemiginowska}


\author[b]{R.M. Suleiman}
\author[a]{M.S. El-Nawawy}
\author[a]{A. Ali}
\affiliation[a]{organization={Astronomy, Space Science and Meteorology Department, Faculty of Science, Cairo University},
             city={Giza},
             country={EGYPT}}

\affiliation[b]{organization={Center for Astrophysics $|$ Harvard \& Smithsonian},
             city={Cambridge},
             postcode={MA 02138},
             country={USA}}

\begin{abstract}
We performed a study of high redshift ($z>2$) quasars, looking for the main differences between Radio Loud Quasars (RLQ) and Radio Quiet Quasars (RQQ) in the X-ray band. 
Our sample of 472 RQQ and 81 RLQ was selected by cross-matching the SDSS DR7 quasars catalog with the Chandra Source Catalog.
We computed the X-ray luminosity for the two samples and
confirmed the X-ray luminosity excess of RLQ over RQQ. 
We fit the X-ray spectra assuming the absorbed power law model and obtained the photon index ($\Gamma$) values for all the sources in the sample.
We excluded quasars with a low number of counts ($<10$) and large uncertainty on the best-fit photon index ($\Gamma_{err}>1$), and obtained the mean values of $\Gamma_{RLQ}=1.70 \hspace{0.5mm}_{-0.33}^{+0.36}$ and $\Gamma_{RQQ}=2.19 \hspace{0.5mm}_{-0.44}^{+0.46}$ for the RLQ and RQQ samples, respectively, showing that the RLQ have flatter (harder) X-ray spectra than RQQ. The Kuiper-two test confirms this result with the significant difference between the RLQ and RQQ photon index distributions ($D_{k}=0.37$ and P-value $= 10^{-6}$). We also evaluated the hardness ratio distributions and confirmed that the spectra of RLQ are flatter than the spectra of the RQQ. The RLQ's hard-to-soft ratio distribution is skewed towards the hard X-ray band, while the RQQ is towards the soft X-ray band. The hard-to-medium and medium-to-soft ratios show no difference. 
\end{abstract}

\begin{keyword}
Radio Loud Quasars\sep Radio Quiet Quasars \sep X-ray Astrophysics\sep X-ray photon index \sep Hardness Ratio
\end{keyword}
\end{frontmatter}

\section{Introduction}  \label{intro}
There are two main classes of quasars: the Radio Quiet Quasars (RQQ) and the Radio Loud Quasars (RLQ). They have been identified based on the orientation and presence of a radio jet \citep{Antonucci1993, WilsonandColbert1995, Urry1995}. RLQ have optical and X-rays luminosity about three times greater than their RQQ counterparts \citep{zamorani1981, Worrall1987, Miller2010, Zhu2020}.
The X-ray radiation could be due to the Compton scattering of UV photons by energetic electrons or due to synchrotron radiation from highly relativistic electrons. 
\citep{Mushotzky1993, Nowak1995, Turner2009, Worrall2009, Fabian2012}.

Majority of quasars are RQQ with the X-ray radiation attributed to a hot corona formed in the accretion flow \citep{HaardtMaraschi1993, Fabian2015, Zhu2020}. 
RLQ are a small minority, about $\approx10\%$ of all quasars, and are characterized by their relativistic jets generated by an accreting supermassive black hole (SMBH) \citep{Padovani2017, Blandford2019}.
The amount of jet radiation contributing to the X-ray spectrum in RLQ is still not fully understood. However, identifying the main radiation components in the X-ray spectrum is important to the estimates of the quasar power.

The RLQ have flatter X-rays spectra (lower photon index value) than those of the RQQ \citep{Reeves1997, Page2005, Miller2010}. The quasar’s hardness ratio is consistent with the spectral slope \citep{Freeman2001, Evans2010, Peca2021}.
The RLQ are divided into Core Dominant (CD) and Lobe Dominant (LD) \citep{HaardtMaraschi1993, WilsonandColbert1995}. The radio emission of CD quasars is dominated by the relativistic jet, while the LD quasars show significant radio emission from the large-scale components in comparison to the core \citep{Falcke1995, Boroson2002}. These two populations might have different X-ray radiation processes, which was noted recently by \cite{Zhu2020}.

During the past decades, quasar data from X-ray surveys have become available, allowing for statistical studies of relatively large samples. 
Many recent studies considered the high redshift quasars for survey \citep{Kelly2007, Vito2019, Pons2020, Lietal2021}. Some studies were focusing on deducing RLQ properties using correlations between X-ray, radio, and optical (or UV) luminosities to investigate the quasars physical model, \cite{Miller2010} investigate the disk-jet model, \cite{Zhu2020} deduced the disk-corona-jet model. 
Interestingly, \cite{Lusso2017} were studying RQQ and showed that RQQ could be used as standard candles at high redshifts ($z>2$), which is important for distance measurement and cosmological tests.

In this research, we investigate the differences in the X-ray spectral properties (photon index, intrinsic absorption, hardness ratios, and X-ray luminosity) between RQQ and RLQ samples using the data available in the Chandra Source Catalog (CSC2) \citep{Evansetal2019}. We study quasars at a high redshift near the peak of cosmic quasar activity, at $z>2$. Our sample contains the largest number of RLQ at high redshift observed with Chandra and include faint sources with [10$^{-15}$ - 10$^{-13}$] erg$\hspace{0.5mm}$cm$^{-2}\hspace{0.5mm}$s$^{-1}$ \footnote{https://cxc.cfa.harvard.edu/csc/char.html} for the energy range [0.5 - 7.0] keV. We calculate the photon index by fitting the faint X-ray spectra, thus expanding the number of quasars with this parameter. The observed Chandra effective energy range is [0.5 - 7.0]\,keV and corresponds to the rest frame energy greater than [1.5 - 21.0]\,keV at redshift $z>2$, so at the higher redshifts we are able to study the X-ray spectra, which are the most sensitive to the properties of the corona and relativistic jet.

In section 2, we briefly describe the data catalogs, the sample selection criteria, and our constraints. In section 3, we show the distributions of RLQ and RQQ as functions of X-ray parameters and illustrate the photon index calculations and constraints, furthermore, we analyze extreme cases for the photon index. In section 4, we discuss our results and compare them with previous studies, and conclude with a discussion and outlook for future work.

\section{Sample selection}  \label{Sample selection}
\begin{figure}[h]
\centering
\includegraphics[width=\linewidth]{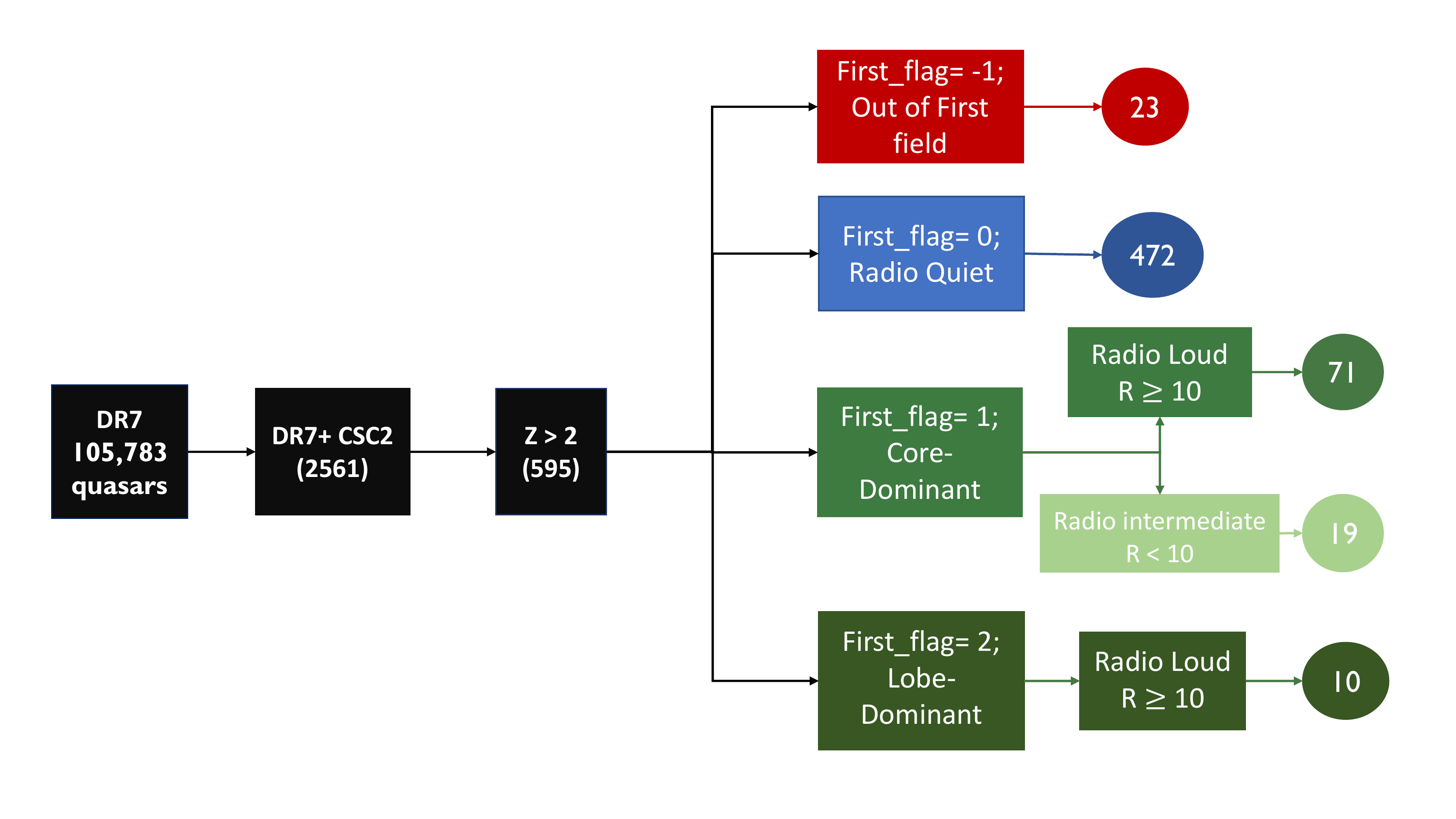} 
\caption{The sample selection is based on the DR7, CSC2, redshift and radio-loudness. The filters in the fourth column are showing the FIRST flag values (-1, 0, 1, and 2) representing quasars (not in the FIRST field, Radio Quiet, Core dominant, and Lobe dominant). The next column filters the radio-loudness ($R$) into RQQ, RLQ, and RIQ. The circles represent the quasar's number in each category.}
\label{chart}
\end{figure}
We study the X-ray properties of quasars using archival data from two quasar catalogs. We use DR7 quasars catalog \citep{Shenetal2011}, which contains 105,783 quasars with optical spectra and redshift measurements. \cite{Shenetal2011} quasars were selected from the SDSS DR7 sample compiled by \cite{Schneider2010} and all have spectroscopic redshift measurements. \cite{Schneider2010} rejected the pipeline redshift measurements for the quasar candidates with images exceeding the PSF size in the r-band. They provide the uncertainty on the redshift measurement to be $_{-}^{+}\hspace{0.5mm}0.004$.

We use the X-ray data obtained by the Chandra X-ray Observatory (Chandra) during the first 15 years of the mission available in the Chandra Source Catalog release 2.0 (CSC2\footnote{https://cxc.cfa.harvard.edu/csc/}). There are more than 315,000 unique X-ray sources in the CSC2 \citep{Evansetal2019}. Chandra has a high-quality angular resolution (better than $\SI{5}{\arcsecond}$), which is important for detecting faint sources, at high redshift, with good source positions.
We cross-matched the 105,783 DR7 quasars with sources in CSC2, using TOPCAT \citep{Taylor2017}, and set a search cone radius of $\SI{30}{\arcsecond}$, consistent with the range of the sources offset uncertainty given by \cite{Evansetal2019}. We found 2,561 sources corresponding to X-ray sources in CSC2.
We study the sources at high redshift ($z>2$). After applying the ($z>2$) filter, we obtained 595 out of 2,561 quasars.
The details of our full sample selection are presented in Fig.\ref{chart}.

\cite{Shenetal2011} matched DR7 optical quasars catalog with Faint Images of the Radio Sky at Twenty Centimeters (FIRST) catalog \citep{White1997}, and estimated the quasar radio loudness parameter ($R$) defining RLQ and RQQ based on the following equation

\begin{equation}\label{eqn:R}
R= \Big(\frac{f_{6\,\rm cm}}{f_{\rm 2500}}\Big) \\
\end{equation}

where $f_{6\,\rm cm}$ and $f_{\rm 2500}$ are the fluxes density ($f_{\nu}$) at rest-frame 6\,cm and 2500\,\AA, respectively.
The flux density in DR7 is determined from the FIRST integrated flux density at 20\,cm assuming a power-law slope of $\alpha_{\nu}= − \,0.5$. The flux density at the rest frame of 2500\,\AA\, is determined by fitting the optical spectrum with a power-law continuum \citep{Shenetal2011}.


Similar to \cite{Jiangetal2007}, \cite{Shenetal2011} have divided RLQ in DR7 into lobe dominant (LD) and core dominant (CD) with FIRST cone radius of $\SI{30}{\arcsecond}$ and $\SI{5}{\arcsecond}$, respectively. 

\cite{Shenetal2011} have removed the effects of galactic extinction in the SDSS spectra using the \cite{Schlegel1998} map, and the Milky Way extinction curve by \cite{Cardelli1989}. Furthermore, \cite{Shenetal2011} shifts the spectra to the rest frame using the cataloged redshift as the systematic redshift \citep{Hewett2010}.

We select the Radio Intermediate quasars (RIQ) to have $R<10$ and RLQ with ($R \geq 10$) \citep{Miller2010}.
We applied the above selection categories to our initial sample of 595 quasars in CSC2 and divided them into different radio-loudness categories as given in Figure \ref{chart}.
Because we focus on strong differences between the RLQ and RQQ, therefore we exclude the intermediate sample and only include RLQ and RQQ in our analysis. Our final sample contains 81 RLQ and 472 RQQ.


\section{Data Analysis and Results}
We study several parameters representing the X-ray properties of the quasars in our samples. The redshift ($z$), and the  radio loudness ($R$) are provided from DR7 \citep{Shenetal2011}, while the X-ray flux ($f_{X}$), the hardness ratios ($H\!R_{h/m}$), the hydrogen column density ($N_{H}$), and the X-ray spectral files are given in CSC2 \citep{Evansetal2019}. We calculate the X-ray luminosity ($L_{X}$) and the X-ray photon index ($\Gamma$). 

In order to evaluate the difference between RLQ and RQQ samples in all X-ray parameters we use the Kuiper-two sample test \citep{Kuiper1960}. The Kuiper test is a test for the difference between two samples based on their observed Cumulative Distribution Functions (CDF). It is an extension of the Kolmogorov–Smirnov test, but it is more sensitive to the shift between the two distributions and the difference in the tails of the distributions. 
The Kuiper test is non-parametric and does not assume any functional form of the sample's true distribution and it is appropriate when true distributions are unknown. The test returns $D_{K}$ and $F_{k}$, which are the maximum difference between the two samples and the probability p-value of the test, respectively. The $F_{k}<0.05$ rejects the hypothesis that the two samples are drawn from the same distribution, so the smaller the value the stronger the significance of the difference between the two samples. All the Kuiper-two test values of this study are given in (Table \ref{tab:Kuiper}).

In our figures, we use normalized density histograms because we have different samples size. The histograms represent the probability density function of the parameter distributions \citep{Hunter2007}, (i.e., $ m / M\times b)$, where $m$ is the number of quasars in each specific bin, $M$ is the total number of quasars in the sample, and $b$ is the bin bandwidth. So the area under the bins integrates into one.
We apply the same number of bins to RLQ and RQQ. The RQQ sample appears to have a smaller bin bandwidth than the RLQ sample because the bin bandwidth is affected by the sample number in the probability density function. 
We also apply the Kernel Density Estimation (KDE) smoothing function to account for the small sample size and different bin sizes \citep{Murray1956}. The small sample size may contribute to the gaps within the histograms, and different binning could lead to statistical biases \citep{Waskom2021}.
We use the following KDE equation:
\begin{equation} \label{eqn:K}
P(x) = \frac{1}{{M \times h}}\sum_{i=1}^{M} {k}\Big(\frac{x-x_{i}}{h}\Big) 
\end{equation} 

Where $M$ is the total number of quasars in the sample,
$h$ the Kernel bandwidth,
$k$ the chosen kernel weight function in our estimate (Gaussian), 
$x$ is the point where to calculate the function, and
$x_{i}$ is the parameter value in bin $i$. The seaborn package \footnote{https://seaborn.pydata.org/generated/seaborn.kdeplot.html} for fitting KDE has a built-in kernel bandwidth optimal estimation using Silverman methods, which are used for random normally distributed samples \citep{Silverman1981}.

Figure \ref{ZX} shows the redshift distributions of RLQ and RQQ samples. We apply the Kuiper-two test which returns a small difference between the RLQ and RQQ samples with $D_{k} = 0.19$ and $F_{k} = 0.08$. This confirms that the RLQ and RQQ samples in our studies have consistent redshift distributions.

\subsection{X-ray Luminosity}
We calculated the X-ray luminosity using the equation given by:

\begin{equation} \label{eqn:X}
L_{X}= 4 \pi  {d_{L}}^{2}  f_{X} 
\end{equation}

Where $L_{X}$ is X-ray luminosity, $d_{L}$ is the distance luminosity, and $f_{X}$ is the X-ray flux in [0.5 - 7.0] keV broadband energy band.
The cosmological model used in this study is the WMAP9 with ($H_{o} = 69.33$,  $\Omega_{o} = 0.287$,  $\Omega_{\Lambda} = 0.712$) parameters \citep{Hinshaw2013}. We use the WMAP9 under the {\tt {astropy.cosmology}} package to obtain the distance luminosity ($d_{L}$) \citep{Astropy2018}.
Using $f_{X}$ and $d_{L}$, and Eq.\ref{eqn:X} we calculate the X-ray luminosity \citep{harris2020array}.

\begin{figure}
    \begin{subfigure}[b]{0.5\textwidth}
        \centering
        \includegraphics[width=\textwidth]{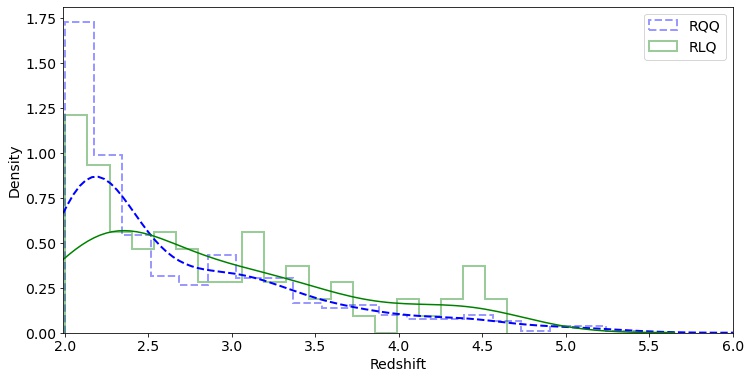}
        \end{subfigure}
    \hfill
    \hfill
    \begin{subfigure}[b]{0.5\textwidth}
        \centering
        \includegraphics[width=\textwidth]{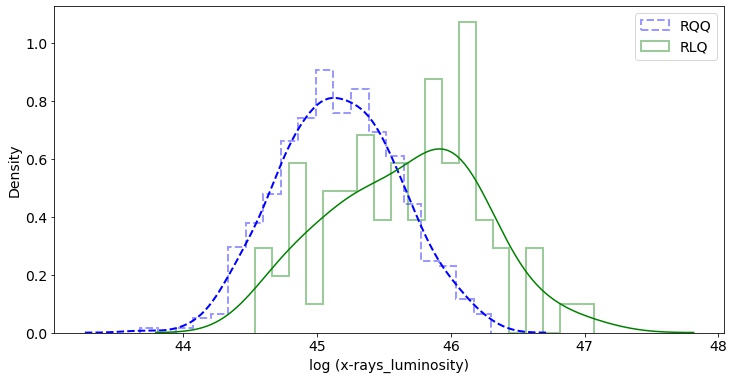}
    \end{subfigure}
\caption{The two panels show the redshift (left), and the X-ray luminosity (right) distributions comparison between RLQ and RQQ. The green histogram represents all of the RLQ as a function of X-ray Luminosity, and the solid green curve represents the KDE for the RLQ. The blue histogram represents the RQQ as a function of the X-ray Luminosity, and the dashed-blue line represents its KDE. The histograms are normalized.}
\label{ZX}
\end{figure}

Figure~\ref{ZX} shows the X-ray luminosity distributions of RLQ and RQQ samples. 
The RQQ KDE (blue) shows a shape consistent with a Gaussian distribution and the RLQ  KDE (green) is skewed to the higher X-ray luminosities. The X-ray luminosity range, given in log scale, for RLQ is $L_{X_{min.}}=44.5$ and $L_{X_{max.}}=47.07$, while for RQQ are $L_{X_{min.}}=43.68$ and $ L_{X_{max.}}=46.30$. The differences between minimum and maximum luminosities are similar, 2.57 and 2.62 for RLQ and RQQ, respectively. However, the median of $L_{X}$ is higher in the RLQ sample by 0.53 compared to the RQQ's median. 
This difference in the median between RLQ and RQQ is significant and indicates a reliable difference between the intrinsic properties of the two samples, RLQ and RQQ. 
The Kuiper-two test returns a significant difference in X-ray luminosity distributions between RLQ and RQQ, $D_{k} = 0.42$, and $F_{k} = 2.18\times10^{-9}$. The $F_{k}$ value validates the remarkable difference in the X-ray luminosity between the radio-quiet and radio-loud quasars (see Table \ref{tab:compare}).

\subsection{The Hardness ratios}
The hardness ratio is defined as the flux ratio between two different Chandra energy bands. The X-ray energy bands in the CSC2 are divided into several categories \footnote{https://cxc.cfa.harvard.edu/csc/columns/ebands.html}:
\begin{itemize}
\label{list:energy}
\item Broad (0.5-7.0) keV
\item Hard (2.0-7.0) keV
\item Medium (1.2-2.0) keV
\item Soft (0.5-1.2) keV
\end{itemize}

\begin{figure}
    \begin{subfigure}[b]{0.3\textwidth}
        \centering
        \includegraphics[width=\textwidth]{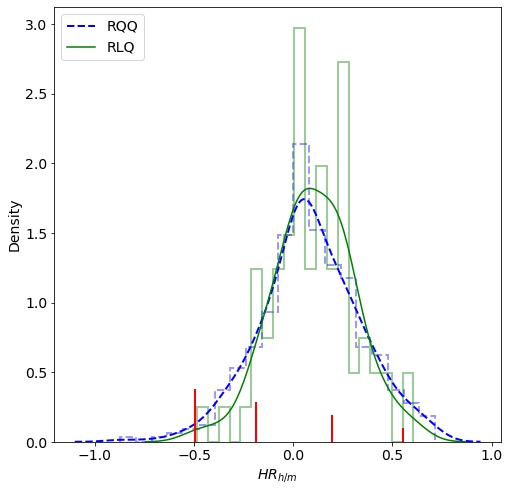}
        \label{HM}
        \end{subfigure}
    \hfill
    \hfill
    \begin{subfigure}[b]{0.3\textwidth}
        \centering
        \includegraphics[width=\textwidth]{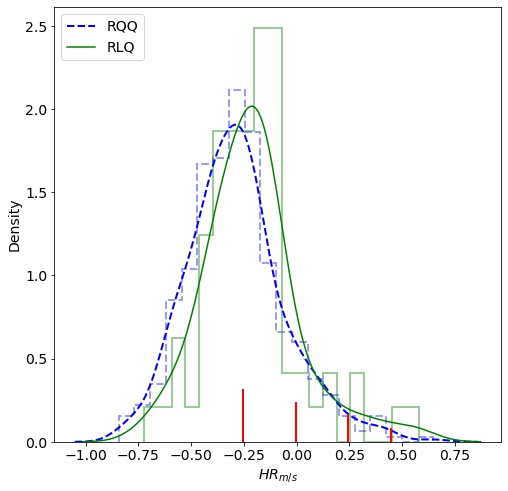}
        \label{MS}
    \end{subfigure}
    \hfill
    \hfill
    \begin{subfigure}[b]{0.3\textwidth}
        \centering
        \includegraphics[width=\textwidth]{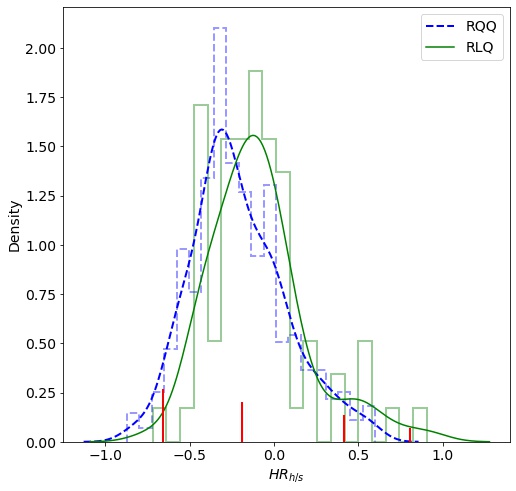}
        \label{SH}
    \end{subfigure}
\caption{The three panels show the $H\!R_{h/m}$, $H\!R_{m/s}$, and $H\!R_{h/s}$ distributions. The green solid line represents the RLQ, while the blue dashed line represents the RQQ. The red vertical lines indicate the photon index values corresponding to each hardness ratio for $\Gamma$ equal to (3, 2, 1, 0) from left to right. The left panel shows no difference between RLQ and RQQ distributions. The middle panel shows a slight difference. The right panel shows a significant difference between RLQ and RQQ with a higher tendency toward soft energy in the RQQ sample.}
\label{HMS}
\end{figure}

The hardness ratios for hard to medium ($H\!R_{h/m}$), medium to soft ($H\!R_{m/s}$), and hard to soft ($H\!R_{h/s}$) \footnote{https://cxc.cfa.harvard.edu/csc/columns/spectral\_properties.html} are given in CSC2 for each detected source.
CSC2 provides $f(h$), $f(m)$, and $f(s)$ the X-ray fluxes in the hard, medium, and soft energy bands, respectively.
\begin{eqnarray}\label{eqn:hard}
    H\!R_{h/m} &=& \frac{f(h)-f(m)}{f(h)+f(m)}  
\end{eqnarray}
The $H\!R_{m/s}$ and $H\!R_{h/s}$ are defined similar to equation \ref{eqn:hard}. When the hardness ratio exceeds zero, the flux of the higher energy band dominates over the flux of the lower energy band. For a general comparison between RLQ and RQQ samples, we investigate the distributions for $H\!R_{h/m}$, $H\!R_{m/s}$ and $H\!R_{h/s}$ shown in Figure~\ref{HMS} and Figure~\ref{HMS}. The distribution plots were normalized and smoothed with KDE.

We also mark the evolution of the photon index as a function of the hardness ratios. Using the {\tt{fake\_pha}} function in Sherpa and the standard ACIS-S response files, we fix the photon index ($\Gamma= 0,1,2,3$) to simulate the spectrum and calculate the corresponding hardness ratios for each $\Gamma$. Figures \ref{HMS} show that the red marks of the photon index decrease (flat spectrum) as the hardness ratios increase (towards the hard band). RLQ and RQQ samples have a similar $H\!R_{h/m}$ distributions (see Figure \ref{HMS}) confirmed by the Kuiper-two test $D_{k} = 0.16$, $F_{k} = 0.37$. 

The $H\!R_{m/s}$ distribution shown in Figure \ref{HMS}, shows a slight shift towards the soft energy band for RQQ in comparison to the RLQ sample, also indicated by the Kuiper-two test $D_{k} = 0.21$ and $F_{k} = 0.05$. 

Finally, the $H\!R_{h/s}$ distribution shows the most significant difference between RLQ and RQQ samples (see Figure \ref{HMS}) with the Kuiper-two test results of $D_{k} = 0.25$ and $F_{k} = 0.01$, the test accuracy is $99.8\%$ (see Table \ref{tab:Kuiper}). The $H\!R_{h/s}$ distribution indicates that the X-ray spectra of RQQ quasars are softer than the spectra of RLQ quasars in our samples.

We investigate the X-ray properties of the CD and LD quasars separately in our comparison to RQQ by applying the Kuiper-two test on all the parameters. We find that LD and CD samples are consistent in all of the X-ray physical parameters except for hardness ratios, $H\!R_{h/s}$ and $H\!R_{h/m}$ with $F_{k} = 0.16$ and $F_{k} = 0.10$, respectively. However, $H\!R_{h/s}$ and $H\!R_{h/m}$ distributions for LD sample are similar to RQQ with $F_{k} = 0.53$ and $F_{k} = 0.58$, respectively. 
On the other hand, our LD sample is small (10 LD quasars). Future studies of CD and LD quasars with high-quality X-ray spectra are needed to confirm and investigate these results further.

\begin{sidewaystable*}
\caption{The Statistical Analysis for X-ray Parameters}
\label{tab:compare}
\small
\centering
\renewcommand{\arraystretch}{2.2}
\begin{threeparttable}
\begin{tabular}{ccccccccccc}
\hline
\multicolumn{1}{c}{} &
\multicolumn{5}{c}{Radio Loud Quasars} &
\multicolumn{5}{c}{Radio Quiet Quasars}\\
\hline
\hline
Parameter& max & min &mean& median & SD & max & min &mean & median & SD \\
\hline
\hline
$z$ &4.7&2.0 &2.88& 2.67& 0.76 &5.42&2.08&2.7&2.45&0.75\\
\hline
$f_{X}$ &10 & 0.05 & 1.4$\hspace{0.5mm}_{-0.2}^{+0.2}$ & 0.7$\hspace{0.5mm}_{-0.14}^{+0.17}$ & 2 & 3 & 0.004 &0.4$\hspace{0.5mm}_{-0.09}^{+0.10}$& 0.2$\hspace{0.5mm}_{-0.68}^{+0.89}$& 0.4\\
\hline
$L_{X}$ &47.07&44.54&45.66$\hspace{0.5mm}_{-0.07}^{+0.08}$&45.9$\hspace{0.5mm}_{-0.07}^{+0.09}$ &0.56 &46.3 &43.68&45.16$\hspace{0.5mm}_{-0.01}^{+0.11}$&45.16$\hspace{0.5mm}_{0.00}^{+0.10}$ &0.45\\
\hline
$H\!R_{h/s}$&0.90&-0.99&-0.10$\hspace{0.5mm}_{-0.35}^{+0.05}$&-0.13$\hspace{0.5mm}_{-0.33}^{+0.06}$&0.3&0.99&-0.99&-0.21$\hspace{0.5mm}_{-0.53}^{-0.05}$&-0.28$\hspace{0.5mm}_{-0.52}^{-0.06}$&0.40\\
\hline
$H\!R_{m/s}$ &0.99&-0.74&-0.21$\hspace{0.5mm}_{-0.39}^{-0.02}$&-0.21$\hspace{0.5mm}_{-0.38}^{+0.05}$&0.29&0.99&-0.99&-0.29$\hspace{0.5mm}_{-0.49}^{-0.02}$&-0.3$\hspace{0.5mm}_{-0.51}^{-0.08}$&0.3 \\
\hline
$H\!R_{h/m}$ &0.6&-0.99&0.11$\hspace{0.5mm}_{-0.16}^{+0.23}$&0.09$\hspace{0.5mm}_{-0.11}^{+0.27}$&0.24 &0.99&-0.99&0.09$\hspace{0.5mm}_{-0.26}^{+0.30}$&0.07$\hspace{0.5mm}_{-0.26}^{+0.28}$& 0.39\\
\hline
$\Gamma$ &3.4&-0.39&1.8$\hspace{0.5mm}_{-0.34}^{+0.38}$&1.76$\hspace{0.5mm}_{-0.32}^{+0.35}$&0.50&4.8&-0.88&2.14$\hspace{0.5mm}_{-0.44}^{+0.0.5}$&2.06$\hspace{0.5mm}_{-0.43}^{+0.48}$&0.65\\
\hline
\hline
\end{tabular}
\begin{tablenotes}
\small 
\item $f_{X}$: The given X-ray flux (erg$\hspace{0.5mm}$cm$^{-2}\hspace{0.5mm}$s$^{-1}$) must be multiplied by factor of $10^{-13}$. 
\item $L_{X}$: The X-ray luminosity (erg$\hspace{0.5mm}$s$^{-1}$) is given in log scale.
\end{tablenotes}
\end{threeparttable}
\end{sidewaystable*}

\subsection{X-ray Spectral Modeling and Photon Index}
CSC2 lists the photon index calculated by fitting a power law model multiplied by the photoelectric absorption. 
However, the CSC2 pipeline restricted the model fitting to spectra with at least 150 net counts (after subtracting the background) and applied the spectral binning of 20 counts per energy bin to use the $\chi^{2}$ fit statistics \citep{Evansetal2019, McCollough2020}. The CSC2 fitting criteria mean that the majority of quasars in our study do not have a photon index available in the CSC2 catalog. We only found 13 RLQ and 26 RQQ. 

On the other hand, CSC2 provides X-ray spectra and response files for all the sources in the catalog. We obtained these spectral files and fit the absorbed power law model to all the quasars in our sample. In order to fit a larger number of quasar's spectra, we put less restrictive criteria. We accept measurements with total counts greater than or equal to 10 and use the wstat-statistics appropriate for low counts data fitting \citep{Freeman2001}. Furthermore, we reject any calculated photon index with an error greater than or equal to one. With these criteria, we increased the number of sources with the calculated photon index, for RQQ from 26 to 455, and RLQ from 13 to 63.

We note that some quasars have multiple observations. In RQQ, there are 85 RQQ quasars with 243 observations. One of these quasars has 11 observations. In the RLQ sample, we found 5 quasars with 12 observations.
We checked for the variability between the multiple observations and confirmed that there is no variability as the measured flux is consistent for each quasar.

\begin{figure}
    \begin{subfigure}[b]{0.5\textwidth}
        \centering
        \includegraphics[width=\textwidth]{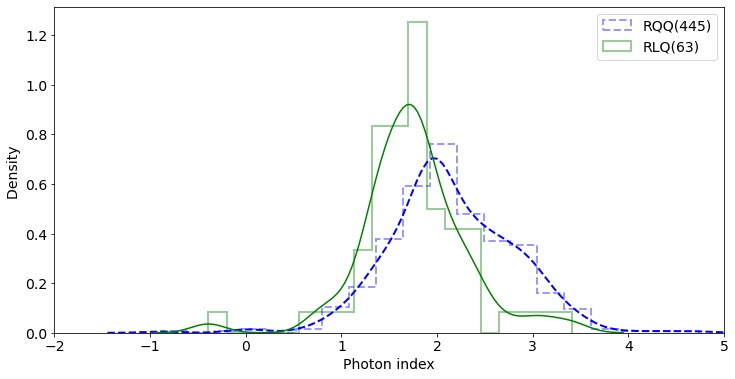}
        \end{subfigure}
    \hfill
    \hfill
    \begin{subfigure}[b]{0.5\textwidth}
        \centering
        \includegraphics[width=\textwidth]{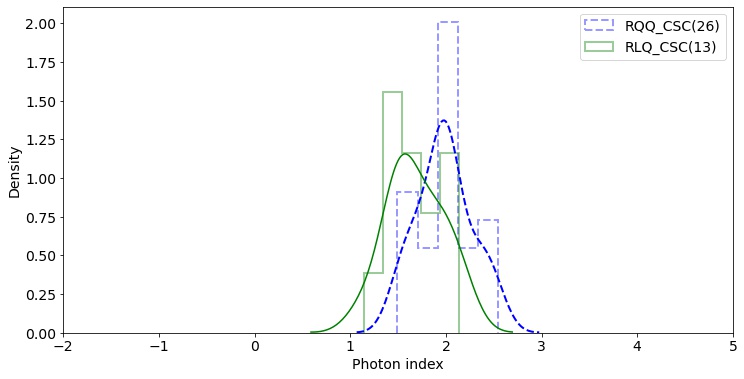}
    \end{subfigure}
\caption{The left panel shows our best-fit photon index using wstat-statistics. We fit 63 RLQ and 445 RQQ. The right panel shows the photon index we obtain from CSC2 for 13 RLQ and 26 RQQ. The RLQ distributions are represented by the solid green histogram and green KDE curve. The RQQ distributions are represented by the dashed-blue histogram and KDE curve. }
\label{fig_GG}
\end{figure}

In Figure~\ref{fig_GG}, 
the left panel shows the distribution of our calculated photon index for the RQQ sample containing 445 quasars and the RLQ sample containing 63 quasars. The right panel shows the CSC2 photon index for the RQQ sample containing 26 quasars and the RLQ sample containing 13 quasars. The fitted photon index has a similar distribution to that of CSC2. We note that a range of the photon index values in our fitting is larger than in the CSC2. We discuss this in Section \ref{extremes}.

The photon index distribution in the RQQ sample shows a steeper spectrum, with the mean value of $\Gamma = 2.14\hspace{0.5mm}_{-0.44}^{+0.05}$, while RLQ shows a flatter spectrum, the mean value of $\Gamma = 1.8\hspace{0.5mm}_{-0.34}^{+0.38}$ (see Table~\ref{tab:Kuiper}). 
In addition, the Kuiper-two test for our calculated photon index shows that the difference between RLQ and RQQ samples is significant with $D_{k} = 0.37$ and $F_{k} = 7.30\times10^{-6}$. However, the Kuiper-two test gives an insignificant difference ($D_{k} = 0.46$ and $F_{k} = 0.18$) for the CSC2 photon index, which may be due to the small sample size (only 13 RLQ and 26 RQQ).  

\subsection{Extreme cases in RLQ and RQQ} \label{extremes}
Figure~\ref{fig_GG} shows some extreme values of the photon index in the distributions (13 RQQ and 3 RLQ). The three RLQ quasars belong to CD class but they show extreme soft spectrum of $\Gamma >3$: i.e. $\Gamma = 3.14\hspace{0.5mm}_{-0.56}^{+0.58}$ , $3.00\hspace{0.5mm}_{-0.56}^{+0.58}$ and $3.4\hspace{0.5mm}_{-0.7}^{+0.8}$ , a total counts $=$ 33, 22 and 13 counts, and the background counts of 0.87, 0.25 and 0.86 counts, at $z =$ 2.23, 2.11 and 3.7, respectively. Due to the high uncertainty of $\Gamma$ and the low number of counts in their spectra we were not able to investigate their properties in more detail. These are interesting outliers identified in our RLQ distribution, which need to be observed in the future.

On the other side, we identify 13 RQQ with extremely hard spectra, $\Gamma <1$ , with a range of the photon index [-0.08 - 0.97]. The total number of counts for these sources range [16 - 827] counts with background counts in the source region [0.12 - 655.8] counts.
We selected three quasars with a relatively good signal-to-noise for detail modeling, with a total number of counts 133, 88, and 156 and a small number of background counts 0.48, 0.27, and 0.88, at $z =$ 2.5, 2.1, and 3.2. These are RQQ with hard spectra potentially indicating a presence of the intrinsic absorption resulting in "flattening" of the intrinsically soft spectrum \citep{Zickgraf1997, de_Kool2002, Page2005}. 

\begin{figure}
    \begin{subfigure}[b]{0.3\textwidth}
        \centering
        \includegraphics[width=\textwidth]{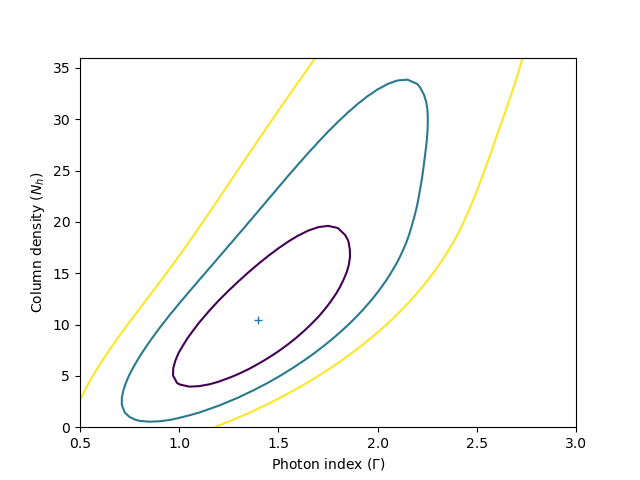}
        \caption{}
        \end{subfigure}
    \hfill
    \begin{subfigure}[b]{0.3\textwidth}
        \centering
        \includegraphics[width=\textwidth]{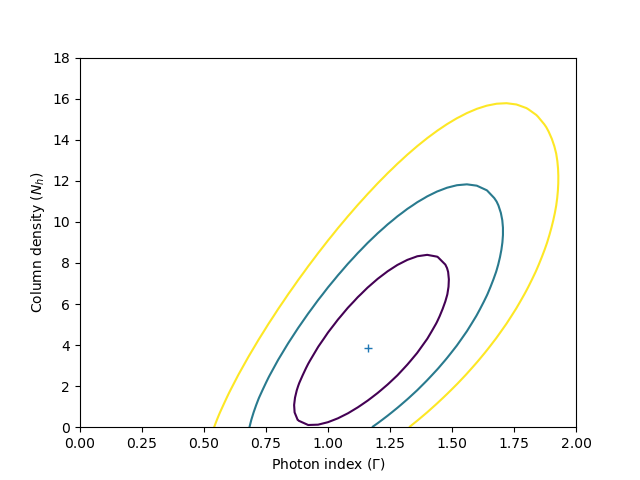}
        \caption{}
    \end{subfigure}
    \hfill
    \begin{subfigure}[b]{0.3\textwidth}
        \centering
        \includegraphics[width=\textwidth]{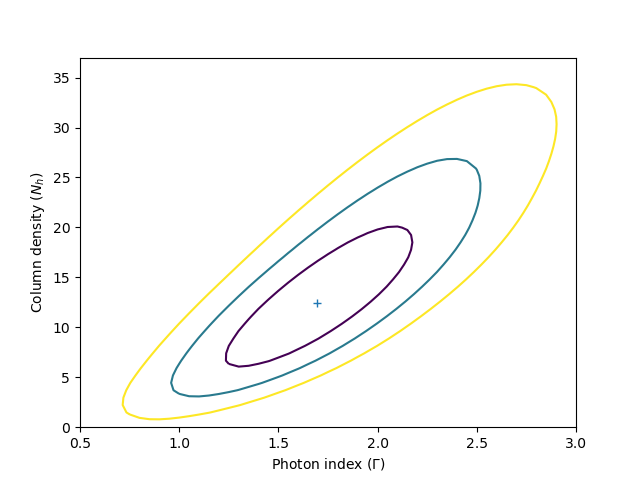}
        \caption{}
    \end{subfigure}
\caption{The confidence regions of $\Gamma$ and $N_{H}$ for the three quasars: (a)2CXO J011513.1+002013, (b)2CXO J123540.1+123620, (c)2CXO J095858.6+020139 with a number of counts (88, 156, 133), respectively. We fit $\Gamma$ and $N_{H}$ with a power law model multiplied by the intrinsic absorption at a given redshift (and including the Galactic absorption). The cross marks the best fit value and the contours show $1\sigma$ (purple), $2\sigma$ (blue) and $3\sigma$ (yellow) levels. The $N_{H}$ values is in log scale.}
\label{fig_reg}
\end{figure}

We fit these three spectra of the RQQ assuming a power law model with additional multiplicative absorption components (Sherpa has built-in models for the intrinsic absorption at the quasar redshift (xszphabs), and the photoelectric Galactic absorption component (xsphabs)). The best-fit $\Gamma$ changes from ($0.80 \hspace{0.5mm}_{-0.14}^{+0.14}$, $0.82 \hspace{0.5mm}_{-0.16}^{+0.16}$, $0.93 \hspace{0.5mm}_{-0.13}^{+0.12}$) to ($1.69 \hspace{0.5mm}_{-0.31}^{+0.31}$, $1.39 \hspace{0.5mm}_{-0.32}^{+0.32}$, $1.16 \hspace{0.5mm}_{-0.21}^{+0.21}$), bringing the photon index values closer to the bulk of the distribution (see Figure~\ref{fig_GG}).
Figure~\ref{fig_reg} shows the confidence contours for the best-fit $\Gamma$ and the intrinsic absorption $N_{H}$ showing a high uncertainty in both the $N_{H}$ and $\Gamma$ values. 
We need higher quality spectra for these quasars to confirm that they are intrinsically absorbed.

\begin{table}[h]
\caption{The Kuiper-two sample test between RLQ and RQQ for all the parameters of interest}
\label{tab:Kuiper}
\centering
\renewcommand{\arraystretch}{1.8}
\setlength{\tabcolsep}{10pt}
\begin{threeparttable}
\begin{tabular}{ccccccccc}
\hline
\multicolumn{1}{c}{Samples } &
\multicolumn{2}{c}{RLQ, RQQ} &
\multicolumn{2}{c}{CD, RQQ} &
\multicolumn{2}{c}{LD, RQQ} &
\multicolumn{2}{c}{CD, LD}
\\
\hline
\hline
Parameters& $D_{k}$ & $F_{k}$ & $D_{k}$ & $F_{k}$ & $D_{k}$ & $F_{k}$&$D_{k}$ & $F_{k}$\\
\hline
\hline
$z$& 0.19 & 0.08 &0.24&0.02 & 0.39& 0.31 &0.50&0.09\\
\hline
$L_{X}$& 0.42 &\textbf{2.18x}$\mathbf{10^{-9}}$& 0.42 &\textbf{2.41x}$\mathbf{10^{-8}}$ & 0.48 &0.09 &0.22&0.99\\
\hline
$\Gamma$ & 0.37 &\textbf{7.30x}$\mathbf{10^{-6}}$&0.39 &\textbf{2.56x}$\mathbf{10^{-5}}$& 0.50 &0.04& 0.24& 0.98\\
\hline
$H\!R_{h/s}$& 0.25 & \textbf{0.01} &0.31 &\textbf{9.80x}$\mathbf{10^{-4}}$&0.37 &0.53& 0.49&0.16\\
\hline
$H\!R_{m/s}$& 0.21 & 0.05 & 0.21 & 0.10 & 0.52& 0.07&0.41&0.37\\
\hline
$H\!R_{h/m}$ & 0.16 & 0.37 & 0.18 & 0.30 &0.34 &0.58&0.52& 0.10\\
\hline
\end{tabular}
\end{threeparttable}
\begin{tablenotes}
\small
\item $D_{k}$: is the maximum absolute difference between the two cumulative distribution functions.
\item $F_{k}$: is a probability (P-value) of the hypothesis that the two samples come from the same population and therefore have the same CDF.
\item Bolded values: are highlighting the highest difference distributions.
\end{tablenotes}
\end{table}

After eliminating the extreme cases, the RLQ $\Gamma_{mean}$ changes from $1.8\hspace{0.5mm}_{-0.34}^{+0.38}$ to $1.70\hspace{0.5mm}_{-0.33}^{+0.36}$ and from $2.14\hspace{0.5mm}_{-0.44}^{+0.0.5}$ to $2.19\hspace{0.5mm}_{-0.44}^{+0.46}$ for RQQ.  
Consequently, the Kuiper-two test value between RLQ and RQQ increased to $D_{k}=0.38$ and its corresponding probability decreased to $F_{k}= 10^{-7}$, which confirms a strong difference between RLQ and RQQ samples.
Since these extreme cases are a small percentage, 4$\%$ RLQ and 2$\%$ RQQ for our sample sets, they are not changing the primary trend of RLQ (hard spectrum) and RQQ (soft spectrum).

\section{Discussion}
We studied a sample of high redshift ($z>2$) quasars selected from CSC2. The samples have similar redshift distribution, but the RQQ sample has (472) a higher number of quasars than the RLQ sample (81). We calculate the X-ray luminosity and the X-ray photon index. All the properties of the two samples are summarized in Table~\ref{tab:compare}.
The Kuiper-two test shows a significant difference between RLQ and RQQ for both $L_{X}$ and $\Gamma$ indicating that the RLQ spectra were flatter than the spectra of RQQ. The Kuiper-two test values for all the X-ray parameters are given in Table~\ref{tab:Kuiper}.

\subsection{Comparing our parameterized results with literature}

Our studies indicate that the X-ray luminosity of RLQ is significantly higher than the X-ray luminosity of RQQ ($D_{k}=0.42$, $F_{k}=2.18\times10^{-9}$), see Table~\ref{tab:compare}) in the sample of $z>2$ quasars in CSC2. This result agrees with the earlier studies \citep{Scott2011}, and suggests an additional X-ray radiation component present in RLQ \citep{Bechtold1994, Zhu2020}.

This additional component may also cause RLQ's X-ray spectra to be flatter than the spectra of RQQ \citep{Reeves2000, Piconcelli2005}. Our studies cover a relatively high rest frame energies, exceeding  30\,keV, in this high redshift sample. These energies are less sensitive to the intrinsic absorption, thus the flattening of the RLQ is less likely related to the absorption (e.g. high absorption columns, $N_H > [10^{22}-10^{26}]$ cm$^{-2}$, are required to modify the high energy spectra), but more likely due to the differences in the radiation processes between the two classes (i.e. RLQ and RQQ).

For our sample, the column density in the direction of the source ranges within [0.57-12.58]$\times10^{20}$\,cm$^{-2}$, with a mean of 2.49$\times10^{20}$\,cm$^{-2}$.
The nuclear obscuration is parameterized by the hydrogen column density $N_{H}$ and the maximum value of $N_{H}$ in our sample is $1.26 \times 10^{21}$\,cm$^{-2}$, which does not affect the AGN X-ray continuum \citep{Hickox2018}.
The obscuration due to the Compton-thick absorption requires a strong reflection component at E $>$ 10 keV, and a prominent Fe-K$\alpha$ emission line at 6.4 keV \citep{Ricci2015}. In our sample spectra, we did not find any Fe-K$\alpha$ emission line.

In addition to the photon index we studied the X-ray hardness ratios for the quasars in the two samples. Our analysis shows, no difference in $H\!R_{h/m}$ between RLQ and RQQ samples, a small difference in $H\!R_{m/s}$, and a moderate difference in $H\!R_{h/s}$ with the RLQ having a harder spectra (see Table~\ref{tab:compare} and Table~\ref{tab:Kuiper}).

The soft X-ray radiation might be produced anywhere in the vicinity of a SMBH in both RLQ and RQQ \citep{Shen2006}. However, we find that the peaks of the $H\!R_{h/s}$ and $H\!R_{m/s}$ distributions (see Figures \ref{HMS}) are shifted towards the soft energy band in RQQ but not in RLQ. Our result indicates that for RQQ, the soft X-ray radiation dominates over the radiation in the hard and medium energy bands. However, for RLQ, the radiation in the hard and medium energy X-ray bands dominates over the soft energy band.

\cite{Page2005} considered a small sample of 7 RQQ and 16 RLQ at ($z>2$) observed with XMM-Newton. They used the broad energy band [0.3 - 10] keV. They found 9 intrinsically absorbed quasars with $N_{H}$ between [1 - 2]$\times 10^{22}$ cm$^{−2}$ in the rest frame of the objects. Using the absorbed power law model, they found that RLQ have flatter spectra than the RQQ counterparts (RLQ $\approx$ 1.55 and RQQ $\approx1.98$). Some studies compare RLQ and RQQ in a specific part of the X-rays (hard band) to specify the corresponding mechanism \citep{Gupta2018, Zhu2020}.
In our study, the RLQ is flatter than RQQ by 0.49 $\hspace{0.5mm}_{-0.11}^{+0.10}$, which is a bigger difference than that found by \cite{Page2005}, due to our larger sample size.
We do not see any clear intrinsically absorbed quasars, which could be due to lower signal-to-noise spectra in our sample. Furthermore, the extreme cases in our samples did not show strong evidence for intrinsic absorption.

\subsection{Calibrating our calculated photon index with CSC2}
\begin{figure}
\centering
\includegraphics[width=\linewidth]{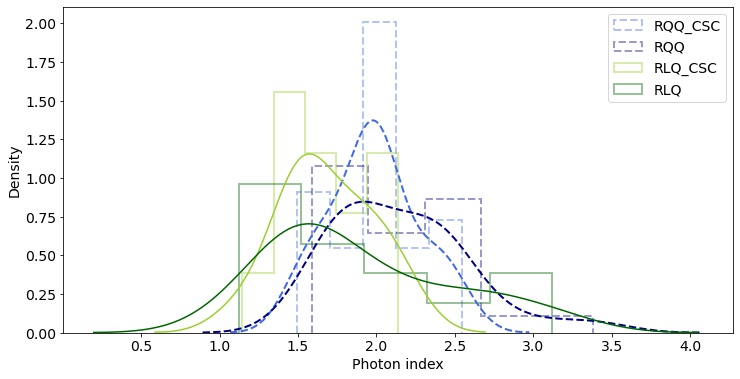} 
\caption{The comparison between our calculated photon index and the CSC2 photon index for the same set of quasars (13 RLQ and 26 RQQ). Dark blue dashed lines and dark green solid lines show our photon index, and light blue dashed lines and light green solid lines show the CSC2 photon index.}
\label{MC}
\end{figure}

\begin{figure}[h]
    \begin{subfigure}[b]{0.49\textwidth}
        \centering
        \includegraphics[width=\textwidth]{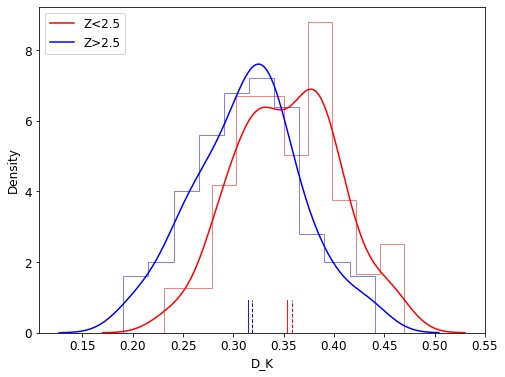}
        \label{nn}
        \end{subfigure}
    \hfill
    \hfill
    \begin{subfigure}[b]{0.49\textwidth}
        \centering
        \includegraphics[width=\textwidth]{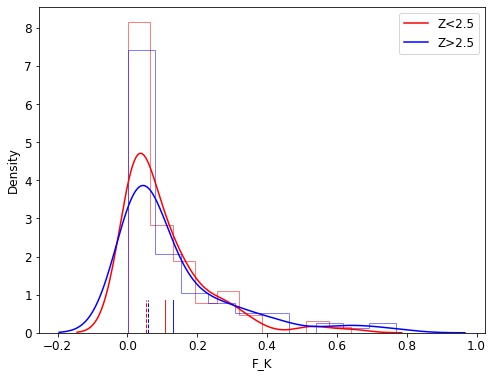}
        \label{mm}
    \end{subfigure}
\caption{The two panels show the 100 trials of the Kuiper test $D_{K}$ (left panel) and $F_{K}$ (right panel) for the $H\!R_{h/s}$ parameter, between RLQ and RQQ. The red color for $H\!R_{h/s}$ at $2<z<2.5$, and the blue color for $H\!R_{h/s}$ at $z>2.5$. The solid vertical lines are for the mean and the dashed vertical lines are for the median.}
\label{MNM}
\end{figure}

We compare the photon index calculated by our spectral modeling to the photon index given in CSC2 for the same quasars (13 RLQ and 26 RQQ). Figure~\ref{MC} shows the RLQ and RQQ distributions for the calculated $\Gamma$ and the one given in CSC2. The distributions show a rough agreement between the two methods, with our modeled values indicating a slightly wider range. 

CSC2 uses the $\chi^{2}$ statistics with background subtraction and binning, while we use wstat-statistics and no background subtraction appropriate for low counts spectra. \cite{vanDyk2001} have explained the $\chi^{2}$ statistical bias at low counts spectra, see also \citep{Protassov2002}. \cite{Humphrey2009} found that even high counts give an inherent bias in the $\chi^{2}$ fitting. These studies show that $\chi^2$ methods should not routinely be used for fitting an arbitrary, parameterized model to Poisson-distributed data, irrespective of the number of counts \citep{Mighell1999}, and instead, the Cash statistic should be adopted \citep{Humphrey2009}.
We used the wstat, which is based on the Poisson likelihood and accounts for the background\footnote{https://cxc.harvard.edu/sherpa/ahelp/wstat.html}.

We applied the Kuiper-two test to evaluate the difference between the two photon-index distributions, $\Gamma_{fit}$ and $\Gamma_{CSC2}$. The test returns high values of $F_{k}$, for RQQ $F_{k} = 0.33$ and $F_{k} = 0.77$ for RLQ, which implies that the distributions of $\Gamma$ resulting from our modeling are consistent with the CSC2 distributions for these small sub-samples.

\subsection{Redshift Dependence of the Hardness Ratio}

Our results on the hardness ratio parameter $H\!R_{h/s}$ indicate that the RQQ spectra are softer than the spectra of RLQ (see Sec.3.2). We perform simulations to confirm that the effect is an inherent physical property of RQQ and is not affected by the redshift. Because the rest frame energy range is shifted towards the lower energy in the observed frame we check the distributions of the hardness ratio parameter in the two redshift ranges. There are 261 RQQ and 33 RLQ at $2<z<2.5$ and 211 RQQ and 48 RLQ at $2.5<z<5.5$ redshift. Thus the fraction of RQQ is higher at $z<2.5$ than at $z>2.5$, which may bias the RQQ's $H\!R_{h/s}$ parameter in the full redshift range.

We apply the Kuiper test to the $H\!R_{h/s}$ at $2<z<2.5$ sample and get the Kuiper parameter values of $D_{K}=0.33$, and $F_{K}=0.04$. Then, we select the same sample size of 33 RQQ and RLQ by randomly selecting 33 quasars from the 261 RQQ sample and using all 33 RLQ in this low redshift bin. 
In the first random selection of the 33 RQQ we get $D_{K}=0.47$, and $F_{k}=0.02$. Afterward, we perform the test for the hardness ratio difference by looking at the distribution of the Kuiper parameters, $D_{k}$ and $F_{k}$, in 100 random samples (see Fig. \ref{MNM}). We selected the 100 random samples of 33 RQQ and used the existing 33 RLQ. The median values for the 100 Kuiper test parameters in this step are $D_{k} = 0.36$, and $F_{K} = 0.05$.

For the $z>2.5$ sample, the Kuiper test results for $H\!R_{h/s}$ difference between RLQ and RQQ are $D_{K} = 0.30$, $F_{K} = 0.04$. We then performed the same simulation steps as described above for the $z>2.5$ sample using the 48 RLQ and a random sample of 48 RQQ selected from the 211 RQQ. The median value of the Kuiper test distribution was $D_{k} = 0.32$, and $F_{K} = 0.06$ (see Fig.\ref{MNM}).

The simulation results show that the difference in the $H\!R_{h/s}$ parameter between RLQ and RQQ samples is slightly more significant at $2<z<2.5$ than $z>2.5$. 
According to \cite{Peca2021}, our sample selection is the least affected by 
the absorption dependence with redshift and the Chandra detector contamination. At low redshift ($z<2$), the difference in the flux between hard and soft bands is larger for quasars with $N_{H}<10^{22}\,\rm cm^{-2}$ because the soft X-ray emission is present in the observed energy band.
At high redshift ($z>2$) the hardness ratio of the quasars with low $N_{H}$ is not affected by the hard band shift to the lower observed energies and only the quasars with high absorption, $N_{H}>10^{23}\,\rm cm^{-2}$, will show the impact on the $H\!R_{h/s}$ parameter. 
We conclude that the observed difference in the hardness ratio between RQQ and RLQ at $z>2$ is not affected by redshift.

\section{Summary and Conclusions}
We studied the X-ray properties of high redshift quasars observed by Chandra. We found a total of 2,561 DR7 quasars in the CSC2 database. After applying redshift and radio-loudness filters we obtained two samples, one with 472 RQQ and the second with 81 RLQ. The two samples have a similar redshift range, $2<z<5$, with the RLQ sample being one of the largest samples of RLQ within that redshift range to date. Our main results are summarized below.

\begin{itemize}    

    \item We found that an average X-ray luminosity of RLQ at high redshift is higher than the average X-ray luminosity of RQQ, consistent with the previous studies.
    
    \item We calculated the mean photon index of $\Gamma_{RLQ}=1.70 \hspace{0.5mm}_{-0.33}^{+0.36}$ and $\Gamma_{RQQ}=2.19 \hspace{0.5mm}_{-0.44}^{+0.46}$ for the RLQ and RQQ samples, respectively. This result confirms that RLQ spectra are flatter than the spectra of RQQ. 
    We identified a few extremely soft RLQ and extremely hard RQQ, but these sources have low signal-to-noise data and require further observations to understand their X-ray properties. 

    \item We found that the LD and RQQ have similar distributions of hardness ratios, $H\!R_{h/m}$ and $H\!R_{h/s}$. In comparison, LD and CD have similar photon index and X-ray luminosity distributions. However, our sample has only 10 LD quasars and more LD observations are needed to confirm this result. 
    
    \item The peaks of $H\!R_{h/s}$ and $H\!R_{m/s}$ distributions are shifted towards negative values (soft energy band) in RQQ compared to RLQ, which confirms that the X-ray luminosity in the RQQ is dominated by soft X-rays in comparison to RLQ.
\end{itemize}   
Our study shows potential directions for further investigation. The quasars of extreme cases need longer observation. The CD and LD comparison needs larger samples for statistically meaningful results. The current samples can be extended to include quasars at higher redshifts, $z>5$, with the future releases of the Chandra Source Catalog. Additionally, the available quasar catalogs can be used to study the early universe population of quasars using high redshift infrared observations which will become available with the JWST \citep{JWST2006}.\\


\textbf{Software}:
Sherpa \citep{Freeman2001}, Topcat \citep{Taylor2017}, Python packages: Astropy \citep{Astropy2018}, Seaborn \citep{Waskom2021}, Numpy \citep{harris2020array}, and Matplotlib \citep{Hunter2007}.\\

\section{Acknowledgement}
This research has made use of data obtained from the Chandra Data Archive and the Chandra Source Catalog, and software provided by the Chandra X-ray Center (CXC) in the application packages CIAO and Sherpa. F.S. thanks CXC Helpdesk and Nick Lee for the support in the analysis of Chandra data. A.S. was supported by NASA contract NAS8-03060 ({\it Chandra} X-ray Center). We are very grateful to the referee for helpful and constructive comments that helped to improve the paper.




\bibliographystyle{elsarticle-harv} 
\bibliography{cas-refs}






\end{document}